\documentclass[useAMS,usenatbib]{mn2e}
\usepackage{graphicx}
\usepackage{latexsym}

\title{A new algorithm for modelling photoionising radiation in smoothed particle hydrodynamics}

\author[J. E. Dale, B. Ercolano, C. J. Clarke]{J. E. Dale$^{1}$ B. Ercolano$^{2,3}$ C. J. Clarke$^{4}$\\
$^{1}$Lund Observatory, Box 43, SE-221 00, Lund, Sweden\\
$^{2}$Harvard-Smithsonian Center for Astrophysics, 60 Garden Street, Cambridge, MA, USA\\
$^{3}$Department of Physics and Astronomy, University College London, 
Gower Street, London WC1E~6BT, UK\\
$^{4}$Institute of Astronomy, Madingley Road, Cambridge, CB3 0HA, UK}
\begin{document}

\pagerange{\pageref{firstpage}--\pageref{lastpage}} \pubyear{2002}

\maketitle

\label{firstpage}

\def\mnras{MNRAS}
\def\apj{ApJ}
\def\aap{A\&A}
\def\apjl{ApJL}
\def\apjs{ApJS}
\def\araa{ARA\&A}

\begin{abstract}
We present a new fast algorithm which allows the simulation of ionising radiation emitted from point sources to be included in high-resolution three-dimensional smoothed particle hydrodynamics simulations of star cluster formation. We employ a Str\"omgren volume technique in which we use the densities of particles near the line-of-sight between the source and a given target particle to locate the ionisation front in the direction of the target. Along with one--dimensional tests, we present fully three--dimensional comparisons of our code with the three--dimensional Monte-Carlo radiative transfer code, {\sc mocassin}, and show that we achieve good agreement, even in the case of highly complex density fields.
\end{abstract}


\section{Introduction} 
\indent Smoothed particle hydrodynamics (SPH) is a powerful and flexible numerical technique which has been applied to numerous astrophysical problems. The wide applicability of SPH derives largely from the fact that it is intrinsically Lagrangian. This allows SPH to deal effortlessly with large density contrasts and obviates the need for a grid or mesh to be imposed on the simulation space, thus placing no restrictions on the symmetry of the problems which may be studied. SPH has been used to model circumstellar and circumbinary disks (\cite{1996MNRAS.279..402M}, \cite{2002pcvr.conf..551S}), stellar collisions (\cite{2002MNRAS.332...49S}, \cite{1996ApJ...468..797L}), the collapse of molecular cores (\cite{1998ApJ...508L..95B}),  the formation of star clusters (\cite{2003MNRAS.339..577B}, \cite{2000ApJ...535..887K}) and galaxy formation (\cite{2004MNRAS.347..740K}). SPH is a proven tool for modelling gas dynamics, and hybridisation with N-body codes allows the inclusion of gravity in SPH simulations, but such comparatively basic input physics limit the sophistication of the simulations that can be performed using SPH. Two problems that have been traditionally very difficult to solve in SPH are the inclusion of magnetic fields and the implementation of radiative transfer. Studies of all the systems mentioned above are likely to benefit considerably from the ability to model these phenomena.\\
\indent In this paper, we describe a new algorithm designed to implement a very simple form of radiative transfer in the context of SPH simulations of the formation of star clusters. Massive stars inject energy and momentum into the ISM by a variety of means and these feedback mechanisms have important consequences on a variety of size scales, from the photoevaporation of circumstellar disks around single or binary stars by ionising radiation (\cite{1996A&A...315..555Y}) to the inflation of galactic superbubbles by the combined action of the radiation, winds and supernova explosions of thousand of O-stars (\cite{2002MNRAS.337.1299C}).\\
\indent On a scale of a few parsecs, midway between the extremes given
above, feedback from massive stars is thought to have a profound
impact on the evolution of star clusters
(e.g. \cite{1984ApJ...285..141L}, \cite{1998orig.conf..150E},
\cite{2003A&A...411..397T}). Since most stars form in clusters
(\cite{2000prpl.conf..151C}), it is clear that the effect of stellar
feedback acting on such scales is a crucial facet of the global star
formation process. It has long been known that our galaxy is forming
stars at only a fraction of the rate which would obtain if star
formation were a process governed solely by gravity
(\cite{1974ApJ...192L.149Z}). It has also become clear that the
majority of star clusters do not survive for long times as shown by (e.g.)
 \cite{2003ARA&A..41...57L} in our Galaxy, \cite{2005A&A...443...79B} in M51 and
 \cite{2006MNRAS.373..752G} in a sample of
 extragalactic young massive clusters. Feedback from
massive stars offers attractive solutions to both of these
problems. The ionising radiation and winds from massive stars drive
powerful outflows into the gas around them. Such outflows can
potentially remove enough mass from a protocluster on short timescales, leaving it unbound; the rapid expulsion of gas inhibits the production of stars, making star-formation a very inefficient process.\\
\indent The influence of stellar feedback on star-formation has been
studied from a variety of theoretical
angles. \cite{1997MNRAS.284..785G},\cite{1980ApJ...235..986H},
\cite{2001MNRAS.323..988G},\cite{2003MNRAS.338..673B} and
\cite{2006MNRAS.373..752G} investigated the effect of removal of the
gaseous content of protoclusters, although they did not model the
mechanism of gas expulsion itself, treating it either as instantaneous
or as occurring on a prescribed timescale. Several studies have
investigated the dispersal of molecular clouds by photoionising
radiation. \cite{1979MNRAS.186...59W} used simple theoretical
arguments to study the erosion of a molecular cloud by O-stars
situated near the edge of the cloud, \cite{1986MNRAS.221..635T} used a
one-dimensional Lagrangian finite-difference code to model the impact
of photoionising radiation on the residual gas in globular clusters
and \cite{1996AAS...188.1708F} conducted two-dimensional grid-based
simulations of the growth of H~{\sc ii} regions. All the theoretical
work described above makes use of some form of symmetry to simplify
the problem. More recently \cite{2004A&A...413..929G} adopted a fractal gas density
distribution to mimic the asymmetries and inhomogeneities of star
clusters. 
However, real protoclusters are inhomogeneous, anisotropic and dynamic environments. The
only way in which to study ionising feedback under realistic conditions is to perform
fully three-dimensional numerical simulations and SPH methods provide
the ideal tool for this task. High resolution SPH simulations are computationally very expensive and the addition of further physics inevitably results in further overheads. However, with the advent of fast parallel machines, such studies have now become technically feasible. \cite{2000MNRAS.315..713K} were the first to attempt this with their studies of radiation-driven implosion of molecular cloud cores (\cite{2003MNRAS.338..545K}).\\
\indent In this paper, we present a fast and robust algorithm designed
to simulate photoionisation in SPH and describe tests performed to
ensure it produces accurate results. In Section 2, we give a brief
overview of the physics of photoionisation. In Section 3 and 4, we
describe how we have implemented the basic physics within the SPH
formalism. In Section 5, we present one--dimensional tests of our
algorithm and compare them to analytic solutions. In Section 6, we
describe the comparison of the SPH algorithm to results generated with
a fully three-dimensional Monte Carlo photoionisation code. Our conclusions are presented in Section 7.\\

\section{The physics of photoionisation}
The idealised problem of the consequences
of the sudden ignition of a source of ionising photons inside a cloud of uniform-density H~{\sc i} has been well studied, first by Str\"omgren \cite{1939ApJ....89..526S} and later by Spitzer \cite{1978ppim.book.....S}.\\
\indent Initially, the ample supply of ionising photons causes an ionisation front (IF) to
progress radially outward from the source at highly-supersonic speed leaving
behind it a spherical H~{\sc ii} region. During this phase of its evolution, the IF is said to be of \textit{R--type}. Geometrical dilution of the radiation field and the recombinations of ions and electrons
within the H~{\sc ii} region progressively reduce the photon flux arriving at the front and slow its progress into the neutral gas. If the neutral atomic gas has number density $n_{0}$,  ionisation of the gas will result in number densities of ions and electrons of $n_{i}$ and $n_{e}$ respectively. If the gas is taken to be H~{\sc i} clearly $n_{i}=n_{e}$ and if the gas is \textit{fully} ionised, then $n_{i}=n_{e}=n_{0}$. The recombination   rate per unit volume is then given by $\alpha n_{i}n_{e}=\alpha n_{0}^{2}$, where $\alpha$ is a recombination coefficient. If the
source's ionising photon flux is $Q_{\mathrm{H}}$ s$^{-1}$ (where any photon whose energy exceeds $13.6$ eV is regarded as ionising), the flux arriving at the ionisation
front at radius $R_{I}$ is given by $Q_{\mathrm{H}}-F$ where $F$ is the Str\"omgren integral given by
\begin{eqnarray} 
F=4\pi \int_{0}^{R_{I}}r^{2}n_{0}^{2}\alpha_{B}dr,
\end{eqnarray} 
where $n_{0}$ has been taken to be constant during the initial phase
of the H~{\sc ii} region's development and
$\alpha_{B}$, taken to be $3.0\times10^{-13}$ cm$^{3}$ s$^{-1}$, is the temperature--dependent `case B' recombination coefficient
which neglects recombinations directly to the hydrogen ground state. The
ionising photons produced by such recombinations are assumed to be absorbed
elsewhere in the H~{\sc ii} region and not to make any contribution to the
ionisation equilibrium. This is commonly referred to as the `on-the-spot' (OTS) 
approximation and is justifiable in cases where the optical depth of the H~{\sc ii} region to secondary ionising photons is smaller than the dimensions of the H~{\sc ii} region. Yorke in \cite{1988STIA...8950499K} shows that, for an O-star with an effective temperature of $40 000$ K, the on-the-spot approximation is valid throughout most of the H~{\sc ii} region at the densities of interest here ($>10^{2}$ cm$^{-3}$).\\ 
\indent As the ionisation front proceeds into the neutral gas surrounding the source, the point is quickly reached where
$Q_{\mathrm{H}}=F$ and the ionisation front can proceed no further. The radius at which this
happens is known as the Str\"omgren radius, $R_{S}$ and is defined by
integration of Equation 1: 
\begin{eqnarray} 
R_{S}=\sqrt[3]{\frac{3Q_{\mathrm{H}}}{4\pi n_{0}^{2}\alpha_{B}}}.
\label{eqn:strom}
\end{eqnarray} 
\indent Since the ionised gas within the H~{\sc ii}
region is expected to reach an equilibrium temperature of $\sim10^{4}$ K,
whereas the ambient temperature of the neutral hydrogen in a typical GMC is $\sim10$ K, a very large  pressure
gradient then exists across the ionisation front. This will cause the
H~{\sc ii} region to expand violently into the surrounding H~{\sc i}, driving a strong shock
before it. This expansion is negligible during most of the initial phase
described above, since the speed with which the ionisation front propagates is
much greater than the speed of sound in the ionised gas ($\sim10$ km s$^{-1}$).
However, once the ionisation front has slowed down to a speed of a few times
the sound speed, the pressure-driven expansion of the H~{\sc ii} region begins
to dominate the evolution of the system. The IF is then said to be \textit{D--type}.\\
\indent By considering
the jump conditions across both the shock front and the ionisation front, \cite{1978ppim.book.....S} derived a simple
expression for the time-dependence of the radius of the ionisation front,
$R_{I}$, during the expansion phase of the H~{\sc ii} region
\begin{eqnarray}
R_{I}(t)=R_{S}\left(1+\frac{7c_{s}t}{4R_{S}}\right)^{\frac{4}{7}},
\label{eqn:spitzer}
\end{eqnarray}
where $c_{s}$ is the isothermal sound speed within the H~{\sc ii} region.\\
\section{Basic Str\"omgren volume method}
We have conducted our simulations using a smoothed particle
hydrodynamics code, fully described in \cite{1995MNRAS.277..362B}. The code is a
hybrid N-body/hydrodynamic code, using a binary tree algorithm for
gravitational calculations, and the SPH formulation to
handle the gas dynamics. Particles are evolved on individual timesteps to improve efficiency. This code is a proven tool for studying
clustered star formation (see, e.g. \cite{2001MNRAS.323..785B}, \cite{2002MNRAS.336..659B}, \cite{2003MNRAS.339..577B}) and is thus ideal for studying the problems outlined
in the Introduction.\\
\indent Stars are represented in the code by sink particles. A sink
particle is a point mass with an accretion radius. Gas particles
straying inside a sink particle's accretion radius are accreted by the
sink particle if they pass a series of the tests, the most obvious of
which is whether or not the gas particle is gravitationally bound to
the sink particle. If the gas particle passes the test, its mass,
momenta and energy are added to those of the sink particle and the gas
particle is removed from the simulation.\\
\indent In previous simulations, the sink particles interact with the
gas particles only via their gravitational fields and the accretion
process described above. We have modified the code to treat one or
more of the sink particles within a simulation as a source of
radiation. The photoionisation algorithm selects a gas particle as a
target and determines whether or not enough ionising photons from a
given radiation source reach the target particle during the current
time-step to ionise it (or, if it is already ionised, to keep it that
way). Consistent with the physical assumptions described above, all gas particles in the simulation are regarded as being fully ionised or fully neutral. Once the ionisation algorithm has decided which particles are ionised, these particles are heated to a temperature of $\sim10^{4}$ K appropriate for the choice of $\alpha_{B}$, and control is returned to the dynamical portion of the code.\\
\indent In a nebula consisting of pure hydrogen and characterised by
conditions where the on-the-spot approximation is valid, secondary
ionising photons do not contribute to the overall ionisation
equilibrium. The photons which do control the ionisation balance are
therefore the primary ionising photons which travel radially outwards
from the radiation source until they are absorbed. Since these photons
follow purely radial trajectories through the gas, the fate of each
photon is independent of those traveling along other radial trajectories. This has the consequence that, in a non-uniform
cloud, each photon can be treated as though it were moving through a
spherically-symmetric cloud possessing whatever radial density profile
exists along the photon's direction of propagation. If the cloud in
question is everywhere sufficiently dense that the on-the-spot
approximation holds, the ionisation structure around a point source of
radiation can then be found by a Str\"omgren volume method in which
the distance from the source to the ionisation front in any given
direction can be found using the radial density profile in that
direction. It is worth noting at this point that in the case of multiple ionising sources
with overlapping Str\"omgren spheres, the algorithm described above
cannot reproduce the radiation field in the overlap region. The
ionised mass will therefore be underestimated by the current
methods. This may have important consequences in the case of crowded
fields and therefore a new version of the photoionisation algorithm
is currently being developed and will be described in a forthcoming
paper.\\
\indent In a spherical cloud with an arbitrary radial density profile
$\rho (r)$, Equation 1 becomes
\begin{eqnarray}
F=\int_{0}^{R_{I}}4\pi r^{2}n(r)^{2}\alpha_{B}dr
\label{eqn:strom_int}
\end{eqnarray}
where the uniform number density $n_{0}$ has been replaced with a
number density $n(r)$ which is a function of radius.\\
\begin{figure}
\includegraphics[width=0.48\textwidth]{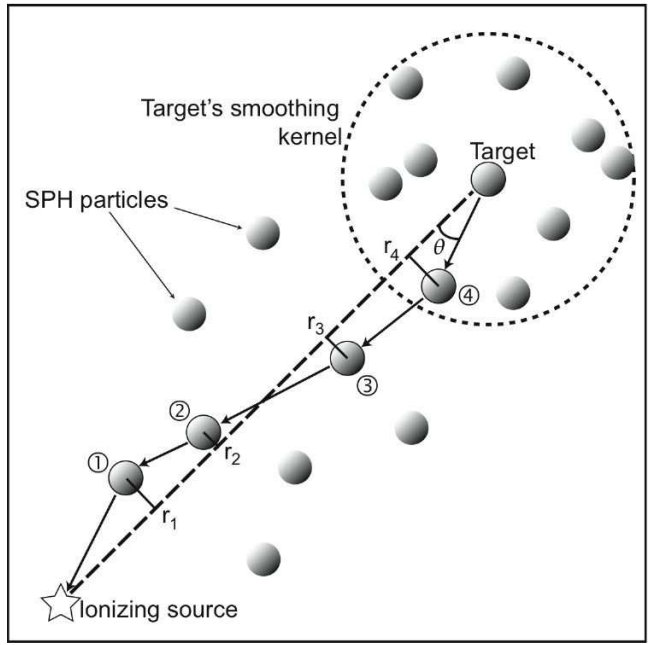}
\caption{Illustration of the method used to select particles whose densities are to be used to derive the density profile along the line-of-sight between the radiation source and a given target particle, denoted by the thick dashed line.}
\label{fig:pathfinder}
\end{figure}
\indent Hence, the flux arriving at any give particle can be found if the function $n(r)$ along the line--of--sight to that particle can be estimated. \cite{2000MNRAS.315..713K} proposed an ingenious method by which the function $n(r)$ can be evaluated in SPH and we make use of it here, with some modifications. The method is illustrated in Figure \ref{fig:pathfinder}. A target SPH particle is first selected and a line-of-sight from this target to the source is drawn.
The target's neighbours (those inside the dotted circle in Figure \ref{fig:pathfinder} are fetched from the
SPH code's neighbour-lists and the one closest to the
line-of-sight is selected (labeled as number 4 in our diagram), that is, the neighbour for which the angle $\theta$
marked in Figure \ref{fig:pathfinder} is the smallest. To save computer time, \cite{2000MNRAS.315..713K} discussed at length the use of a tolerance angle $\Theta$, such that the first particle in a given target's neighbour list which satisfies $\theta<\Theta$ is selected, but we found that the performance of our algorithm remains acceptable even when it is forced to search for the \textit{closest} particle to the line of sight. The position of the selected particle is then projected onto the line-of-sight to produce what \cite{2000MNRAS.315..713K} term an \textit{evaluation point}, labelled $r_{4}$ in Figure \ref{fig:pathfinder}. The neighbours of the newly-chosen particle are then fetched and the selection process is repeated until the radiation source is reached. Once this process has been completed, \cite{2000MNRAS.315..713K} then solve an equation for the rate of ionisation of the target particle. Instead, we use the density profile determined by the process outlined above in a discretised version of Equation \ref{eqn:strom_int} to decide whether the target particle receives enough radiation flux to ionise it, a faster and more robust method than that employed by \cite{2000MNRAS.315..713K}.\\
\indent From the densities of the $N_{LOS}$ particles selected along the
line-of-sight our target, we know the value of $n(r)$ in a
series of $N_{LOS}$ radial bins whose inner and outer radii are defined by the
projected positions of the particles along the line. We assume that the densities at these projected positions are simply the same as the densities of the relevant particle $i$, evaluated in the usual way as
\begin{eqnarray}
\rho_{i}=m_{i}W(0,h_{i})+\sum_{j=1}^{N_{neigh}}m_{j}W(r_{ij},h_{ij}),
\end{eqnarray}
where $m$ denotes the mass of each particle and the sum runs over the $N_{neigh}$ neighbours of particle $i$ (defined as all other particles within two smoothing lengths, $2h_{i}$, of particle $i$). $W$ is the smoothing kernel, a function of the interparticle separations $r_{ij}$ and the average smoothing lengths $h_{ij}$ of particle $i$ and each of its neighbours.\\
\indent Alternatively, we could estimate the densities at the evaluation points by performing an SPH density estimation of the form 
\begin{eqnarray}
\rho_{k}=\sum_{j=1}^{N}m_{j}W(r_{jk},h_{j}),
\label{eqn:eval_point}
\end{eqnarray}
where the sum instead runs over all $N$ SPH particles that overlap the $k$--th evaluation point, $r_{jk}$ is the distance of each of these particles from the evaluation point and $h_{j}$ is the individual smoothing length of each particle. However, identifying the $N$ particles is not trivial; some \textit{but not all} are likely to be neighbours of the $k$--th particle on the line--of--sight  to which the evaluation point belongs. Finding them is therefore time--consuming and we find that in any case it makes very little difference to the results;  the density of the $k$--th particle is almost always a good estimator of the density at the $k$--th evaluation point.\\
\indent We now approximate the integral in Equation \ref{eqn:strom_int} by a sum as follows:
\begin{eqnarray}
\frac{F}{4\pi} = \sum_{i=1}^{N_{LOS}} r_{i}^{2}\langle n(r_{i})\rangle^{2}\alpha_{B}\Delta r_{i}.
\label{eqn:strom_sum}
\end{eqnarray}
The inner radius of bin $i$ is
defined by the radial position of particle $i-1$ projected onto the line--of--sight, $r(i-1)$, and the outer radius of bin $i$ by the
projected radial position of particle $i$, $r(i)$. The width of each bin is then $r(i) -
r(i-1) = \Delta r_{i}$. $\langle n(r_{i})\rangle$ is the average number density of the two SPH particles whose positions define the bin, $[n(i)+n(i-1)]/2$. The density of the point mass, used in calculating
the number density of the first bin, is taken to be zero. Once the sum
has been calculated, $F$ can be subtracted from the source luminosity $Q_{\mathrm{H}}$ to obtain the flux of
photons arriving at the target particle. We can then determine the particle's ionisation state during the current time-step.\\
\section{Improvements over the basic Str\"omgren volume method}
Although the method described above is adequate for the simulation of the growth of a spherical H~{\sc ii} region in a uniform motionless cloud, our code is also intended for use in highly--inhomogeneous and dynamic situations. It is highly likely under such circumstances that neutral gas may enter the H~{\sc ii} region from outside, in an accretion flow, for example, and that ionised gas may either leave the H~{\sc ii} region, or be otherwise cut off from the supply of photons required to prevent it recombining. We have therefore improved upon the basic Str\"omgren volume method to cope with these eventualities.\\ 
\subsection{Neutral gas inside the H~{\sc ii} region}
Modifying the method above to account for the photons required to ionise neutral material which has entered the H~{\sc ii} region (perhaps via an accretion flow), or to grow the H~{\sc ii} region from scratch (the R--type phase of the IF's evolution) is not difficult. If we introduce an ionisation fraction $x_{i}$ (still taken to be either $0$ or $1$) for each particle, then we can rewrite Equation \ref{eqn:strom_sum} as
\begin{eqnarray}
\frac{F}{4\pi} = \sum_{i=1}^{N_{LOS}} r_{i}^{2}\langle n(r_{i})\rangle^{2}\alpha_{B}\Delta r_{i}+\sum_{i=1}^{N_{LOS}} r_{i}^{2}\langle n(r_{i})\rangle (1-x_{i})\frac{\Delta r_{i}}{\Delta t}.
\end{eqnarray}
The first term in the equation, the flux required to keep pace with the recombinations in ionised gas, is always present. The second term allows for the possibility that some flux may be consumed by ionising the gas from a neutral state first, and makes the expression time--dependent. The term $\Delta r_{i}$/$\Delta t$ is the discretized speed at which the ionisation front propagates. The timestep on which the algorithm evolves the ionisation front, $\Delta t$, is usually set to the shortest dynamical timestep in use by the SPH code at the time, so that the H~{\sc ii} region is updated whenever \textit{any} SPH particles are updated. If this timestep is very short, the flux arriving at a given target particle may not be sufficient to ionise it immediately, in which case the flux is banked -- allowed to accumulate until the particle can be ionised. Particles with banked flux are not heated above the neutral gas temperature.\\
\subsection{Ionised gas straying outside the H~{\sc ii} region}
Ionised material that is for some reason deprived of photons will
recombine and eventually cool. To allow for this eventuality, ionised
particles which are receiving no ionising flux are flagged and
monitored by a neutralisation routine. This routine reduces the
ionisation fraction of such particles according to their local
recombination timescale, $(\alpha n)^{-1}$. If the ionisation fraction
of a particle drops below a half, it is assumed to become fully
neutral. \\
\indent \cite{2000MNRAS.315..713K} note that cooling recently-recombined gas instantaneously back down to the original ambient gas temperature is not a good treatment of the physics of recombination, since recombined gas, although neutral, is likely to still be hot. Neutralised particles are therefore passed on to a cooling routine which assigns them new temperatures according to the cooling curve described by \cite{1993A&A...273..318S} until they reach a temperature of $100$ K, at which point they are returned to the initial average gas temperature (usually $10$ K).\\
\subsection{The ionisation code's decision tree}
\begin{figure}
\includegraphics[width=0.45\textwidth]{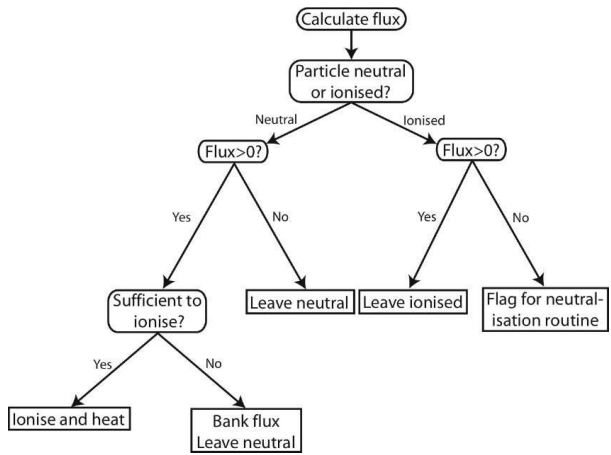}
\caption{Decision tree for the SPH ionisation code.}
\label{fig:tree}
\end{figure}

\indent Once the flux reaching a given target particle has been determined, taking into account the consumption of photons by neutral material lying on the line--of--sight from the source, the ionisation state of the target particle is determined using the decision tree illustrated in Figure \ref{fig:tree}. Note that in this tree, the approximation is made that if the flux reaching an \textit{ionised} particle is non--zero, that particle remains ionised. It could be argued that the flux must in fact be sufficient to exceed the recombination rate in the particle. However, in all cases, the fluxes calculated by the code are those reaching the \textit{centres} of particles and the decrement in flux incurred between the particle edge and its centre is taken account of. The only particles for which this assumption is questionable are those at the ionisation front. For a given ionised particle $i$ at the ionisation front, a non--zero flux reaches the centre of particle $i$ but zero or `negative' flux reaches the next particle further out along the line--of--sight to the source, $j$. This means that the true ionisation front lies somewhere between the centres of $i$ and $j$ and thus somewhere \textit{inside} both particles (recall that SPH particles overlap one another). The decision to keep $i$ ionised and $j$ neutral introduces an error in the location of the ionisation front along the line--of--sight on which $i$ and $j$ lie of $\sim h_{i}$, consistent with the accuracy to which any spatial quantity can be meaningfully determined in SPH.\\

\section{One dimensional tests}
The most basic tests one can perform with our hybrid hydrodynamics/ionisation code are to follow the approach to the Str\"omgren radius and the subsequent expansion of the H~{\sc ii} region.\\
\indent The sudden ignition of a source of $Q_{\mathrm{H}}$ ionising photons per second in a uniform cloud of atomic number density $n_{0}$ leads to the highly supersonic expansion of an R--type ionisation front. The expansion velocity of the front can be derived by considering the rate of fresh ionisations at the front. If the IF has an instantaneous radius $R_{I}$ and the flux reaching the front is $F_{I}$, we may write
\begin{eqnarray}
\frac{dR_{I}}{dt}=\frac{F_{I}}{4\pi n_{0}R_{I}^{2}}=\frac{1}{4\pi n_{0}R_{I}^{2}}\left(Q_{\mathrm{H}}-\frac{4}{3}\pi n_{0}^{2}R_{I}^{2}\alpha_{B}\right).
\end{eqnarray} 
This can easily be integrated and combined with Equation \ref{eqn:strom} to yield
\begin{eqnarray}
R_{I}(t)=R_{S}\left(1-\textrm{exp}\left[-n_{0}\alpha_{B}t\right]\right)^{\frac{1}{3}}.
\label{eqn:strom_approach}
\end{eqnarray}

\begin{figure}
\includegraphics[width=0.45\textwidth]{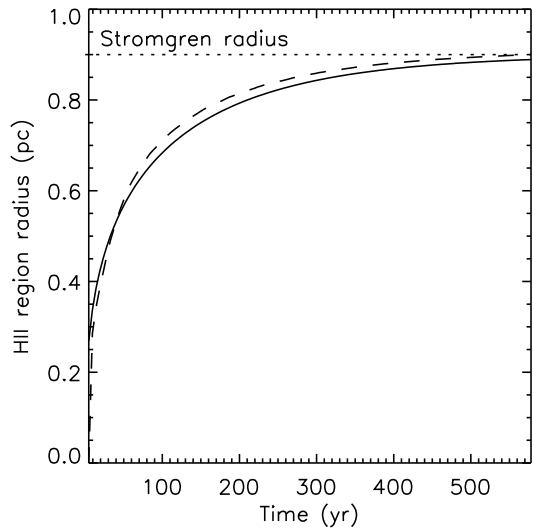}
\caption{Evolution of the radius of the ionisation front with no heating of ionised particles, comparing the approach to the Str\"omgren radius produced by the SPH ionisation code (dashed line) to the analytical solution given in Equation \ref{eqn:strom_approach} (solid line). The Str\"omgren radius itself is shown as a dotted line.}
\label{fig:strom_approach}
\end{figure}

In Figure \ref{fig:strom_approach} we show the results of the ignition of an ionising source in a uniform cloud, following the approach of the ionisation front to the Str\"omgren radius. To simulate this, the heating of ionised particles was disabled and the timestep on which the code dumps output was set to a very small value. We see that the agreement between the SPH results and the analytic solution is excellent.\\
\indent As the ionisation front approaches the Str\"omgren radius, its velocity slows towards the speed of sound in the neutral material surrounding the H~{\sc ii} region. The IF then transitions to its D--type phase according to Equation \ref{eqn:spitzer} (\cite{1978ppim.book.....S}). To test the ability of our code to replicate the Spitzer solution at different numerical resolutions, we constructed three spherical clouds of identical mass $M_{cld}$ and radius $R_{cld}$, containing ionising sources of identical luminosities $Q_{\mathrm{H}}$, but built form different numbers of particles. Since the Str\"omgren radius in the three clouds is the same, the initial H~{\sc ii} regions in the clouds contain different numbers of particles. We chose $M_{cld}=1700$M$_{\odot}$, $R_{cld}=1$pc, $Q_{\mathrm{H}}=10^{49}$s$^{-1}$, so that $n_{0}=1.65\times10^{4}$cm$^{-3}$ and $R_{S}=0.1$pc. We chose particle numbers of $10^{6}$ (run HR), $10^{5}$ (run MR) and $10^{4}$ (run LR), so that the initial Str\"omgren radii contain respectively $1000$, $100$ and $10$ particles.\\
\indent It is not \textit{a priori} obvious what the minimum
requirement to resolve an H~{\sc ii} region in SPH might be. To
resolve an object in SPH is usually taken to require at least the
average number of particles contained within the SPH smoothing kernel
(usually $50$). \cite{1997MNRAS.288.1060B} found that, to resolve
gravitational fragmentation correctly required that a Jeans mass
contain at least $100$ particles (twice the canonical resolution
limit). However, \cite{2006A&A...450..881H} find that even seriously
under--resolving the Jeans mass (with $<10$ particles) does not result
in spurious fragmentation. Our run LR also under--resolves the H~{\sc
  ii} region, in the sense that ten particles is well below the
canonical SPH resolution limit, meaning that the H~{\sc ii} region may
behave unpredictably.\\
\begin{figure}
\includegraphics[width=0.5\textwidth]{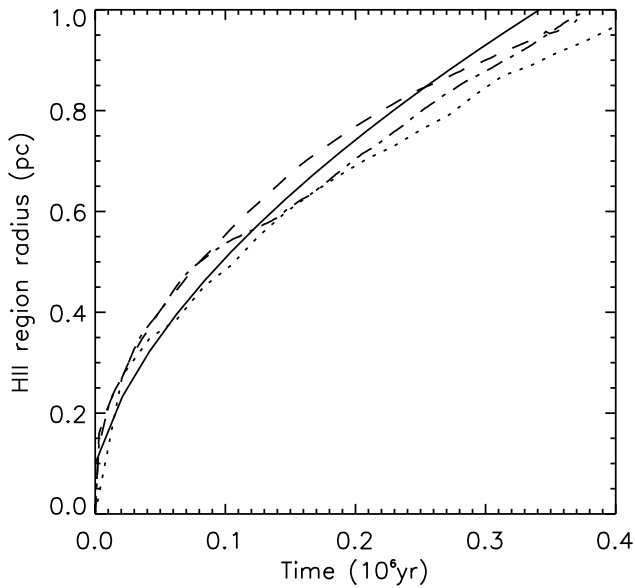}
\caption{Evolution of the radius of the ionisation front in the LR
  (dotted line), MR (dashed line) and HR (dash--dot line)
  calculations, compared with the Spitzer solution (solid line).}
\label{fig:hii_compare}
\end{figure}
\indent In Figure \ref{fig:hii_compare}, we plot the time--evolution of the ionisation front radius in our three models together with the Spitzer solution for comparison. We see that the three SPH models agree very well both with each other and with the Spitzer solution, at least for the evolution of the H~{\sc ii} region from $R_{S}$ to $\sim10R_{S}$. Even the low--resolution calculation gives reasonable agreement with the analytical solution over the range of radii considered. For this simple problem at least, our implementation can faithfully follow the evolution of an H~{\sc ii} region which would be considered severely under--resolved by the usual SPH resolution criteria.\\

\section{Comparison with a 3D Monte Carlo photoionisation code}
Since we intend to use our ionisation code in situations where the gas
density distribution is highly inhomogeneous, one--dimensional tests can
only take us so far. In order to perform a more stringent test than
those presented in the previous section, we therefore selected a very
complex density distribution derived from a three--dimensional
simulation of clustered star formation, and compared the H~{\sc ii}
regions generated by our code and by the fully 3D Monte Carlo
photoionisation and radiative transfer code, {\sc mocassin}
\citep{2003MNRAS.340.1136E, 2005MNRAS.362.1038E}. This code adopts a
stochastic approach to the transfer of radiation, by simulating the
individual processes of absorption and re-emission of radiation quanta
as they leave an ionisation source and diffuse through a gas. The
polychromatic nature of both the primary (stellar) and secondary
(diffuse) components of the radiation field are accounted for
self-consistently as the photon trajectories are computed and the
ionisation and thermal structure of the nebula established. All
relevant atomic physics processes are included and the atomic
database is updated regularly, currently using a combination the
Chianti V5 \citep{2006ApJS..162..261L} database for collision strengths, transition
probabilities and energy levels of heavy metals, the opacity project
for photoionisation cross-section and a new calculation of the H$^+$,
He$^+$ and He$^{2+}$ free-bound continua presented by
\cite{2006MNRAS.372.1875E}.\\

\indent Dust grains mixed with gas inside the ionised regions may also affect
the global ionisation structure of the gas. The grains compete with
the gas for the absorption of UV radiation, hence reducing the flux
available for the photoionisation of atom and ions. This can be
self-consistently accounted for by {\sc mocassin}
\citep{2005MNRAS.362.1038E}, but not by the SPH ionisation
algorithm, since the amount of radiation subtracted from the field, and thus
available to the gas, depends heavily on a number of variables,
including the local dust-to-gas ratio, the grain chemistry and size
distribution, as well as the shape of the {\it local} radiation field,
which cannot be obtained by the SPH ionisation algorithm. In order to be able to compare the outputs from the two
codes we did not include any dust in the {\sc mocassin} calculation,
nevertheless we point out that the ionised masses thus obtained may be
slightly overestimated. \\

\begin{figure*}
\includegraphics[width=\textwidth]{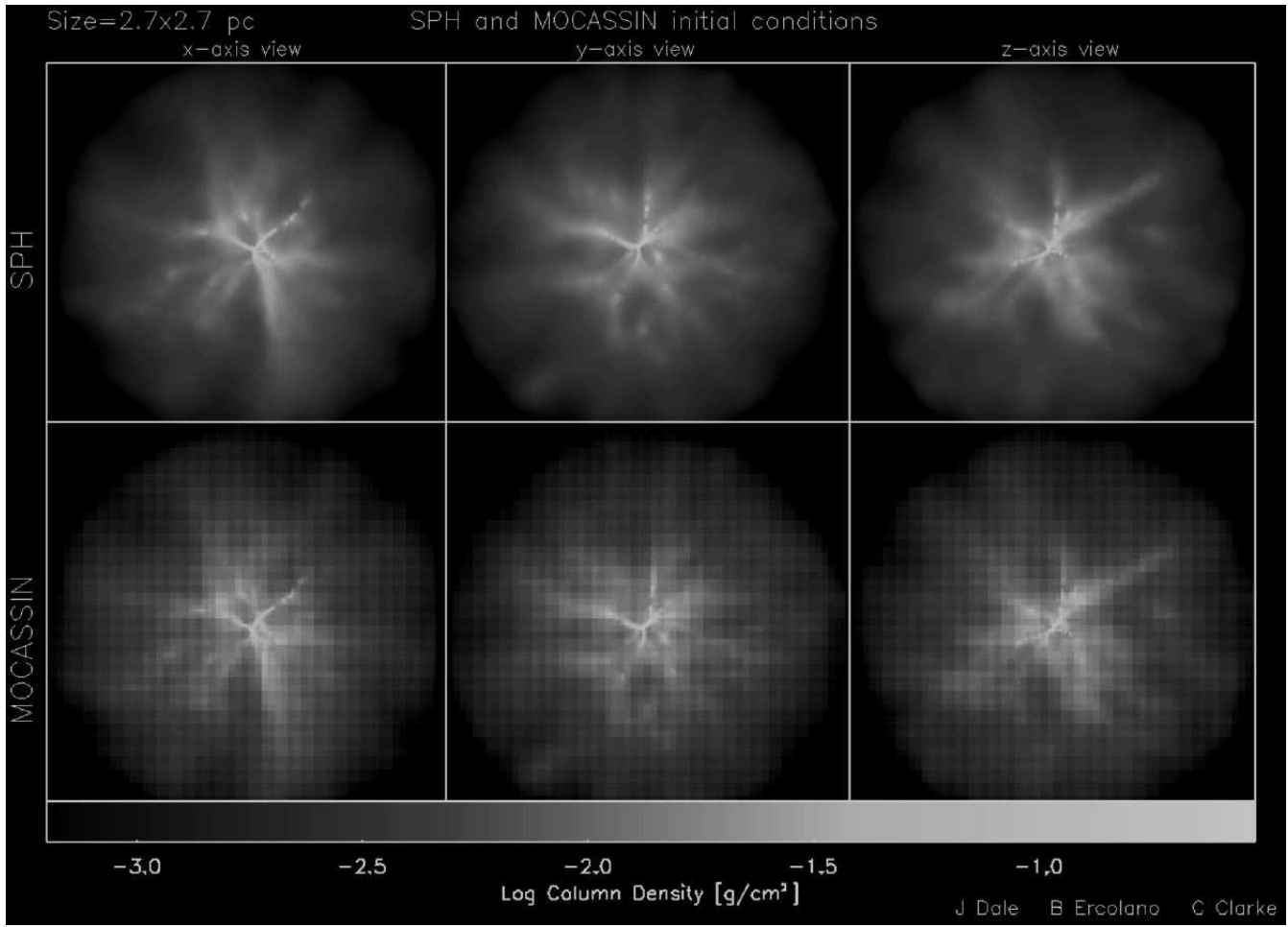}
\caption{Comparison the initial conditions given to the SPH ionisation code (top row) and the {\sc mocassin} radiative transfer code of Ercolano et al (bottom row). Shown are column density maps of the gas distribution fed to  the two codes as viewed along the $x$-- (leftmost two panels), $y$-- (centre two panels) and $z$--axes (rightmost two panels). The ionising source is positioned at (0,0,0), at the centre of all the images.}
\label{fig:grid_allgas}
\end{figure*}

\begin{figure*}
\includegraphics[width=\textwidth]{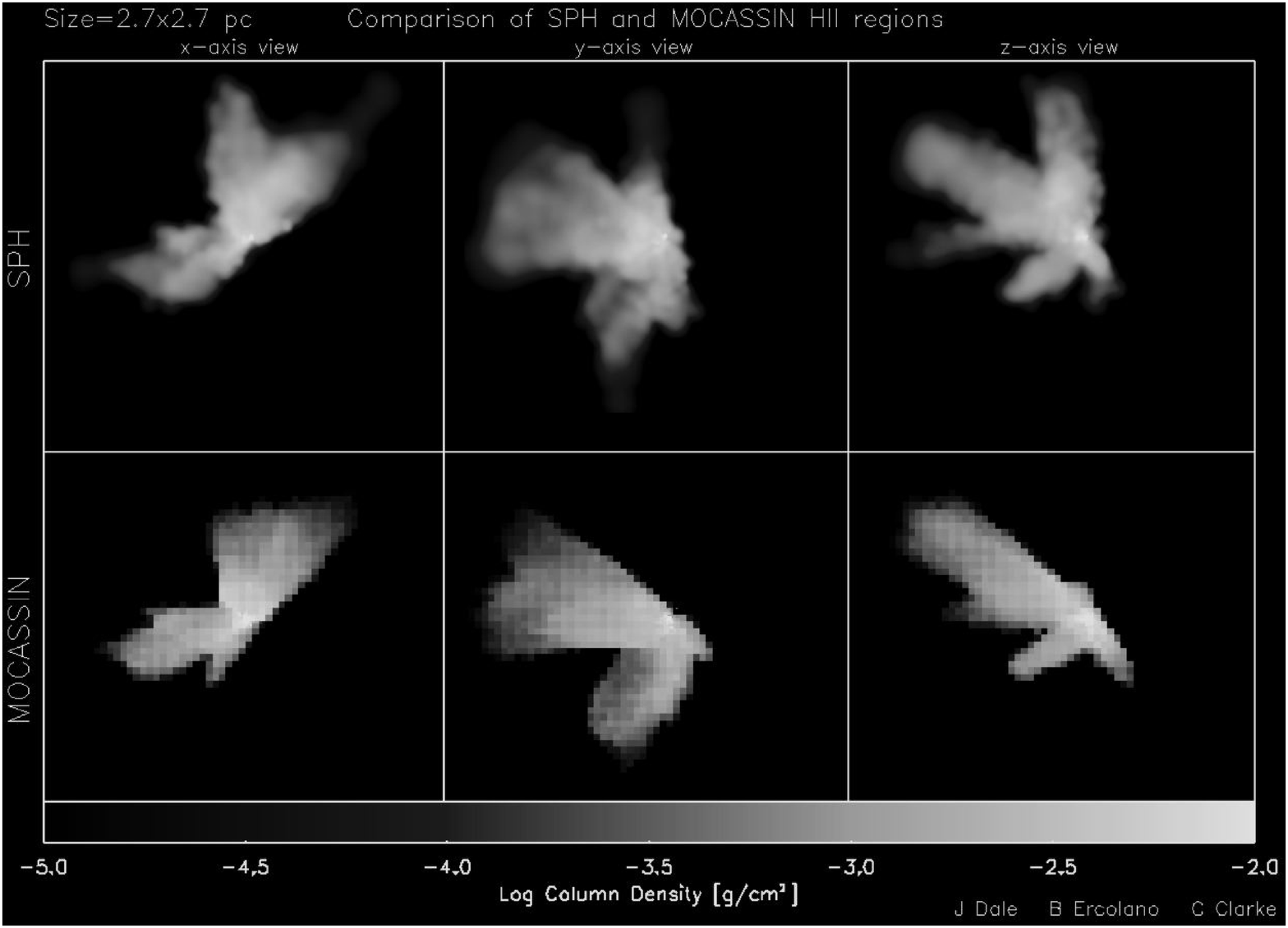}
\caption{Comparison of H~{\sc ii} regions generated by the SPH ionisation code and the {\sc mocassin} radiative transfer code of Ercolano et al. Shown are column density maps of the H~{\sc ii} region produced by the two codes as viewed along the $x$-- (left two panels), $y$-- (centre two panels) and $z$--axes (right two panels).}
\label{fig:grid_hii}
\end{figure*}

\indent We chose to compare {\sc {\sc mocassin}} and the SPH ionisation algorithm under realistic astrophysical conditions. The gas distribution selected for the comparison was drawn from an SPH simulation of the collapse of a molecular cloud originally performed by \cite{2002MNRAS.336..659B} and repeated by \cite{2005MNRAS.358..291D}, who modeled the effect on the cloud and the stellar cluster formed by it of the ionising radiation from the single O--star born at the cluster centre (using the code described here). Three views of the gas distribution are shown in the top panels of Figure \ref{fig:grid_allgas}. The ionising source is at the centre of the image, at the intersection of several dense filaments of gas. The density structure of the cloud is evidently highly inhomogeneous -- maximum and minimum densities are $\sim2\times10^{7}$ and $\sim2\times10^{1}$ cm$^{-3}$ respectively, with mean and median densities of $\sim2\times10^{4}$ and $\sim4\times10^{2}$ cm$^{-3}$ respectively (these densities are derived from the densities of the SPH particles). Note that, as well as exhibiting a very large dynamic range, these densities are sufficiently high that the OTS approximation will hold. Modelling photoionisation in this highly inhomogeneous (and, from the source's point of view, highly anisotropic) environment is clearly very challenging.\\
\indent {\sc mocassin} is a grid--based code, in which a
`mother--grid' of arbitrary resolution is defined; any cell in this
grid may be further divided into a finer subgrid of arbitrary
resolution, so that arbitrary gas distributions can be efficiently
represented in the code with adequate detail. SPH is a gridless
method, representing gas distributions by an ensemble of overlapping
particles of (usually) equal mass, but varying radius (the smoothing
length). We therefore devised a method to interpolate the SPH gas
distribution onto a grid suitable for use by {\sc mocassin} and
subdivided in such a way that the resolution of the grid was
everywhere as close as possible to the resolution of the SPH particle
ensemble. We first generated a uniformly spaced $63^3$ mother grid with cell size $d_{max}$ filling the same volume as the SPH particle distribution. Each SPH particle was then located in the grid and the smoothing length of the smallest particle, $h_{min},$ (if any) in each mother grid cell determined. If for any cell $d_{max}>2h_{min}$, the cell is subdivided into $\lbrace1+\mathrm{INT}[d_{max}/(2h_{min})]\rbrace^{3}$ cells. Finer subdivision of grid cells, for example by $h_{min}$ or $h_{min}/2$ did not result in a significant improvement in the interpolation accuracy. This procedure resulted in the division of 223 of the 250 047 mother grid cells into subgrids. Once the simulation volume is suitably partitioned, the mass of each cell is calculated by finding all $N_{cell}$ SPH particles overlapping the centre of the cell, estimating the density at that point using Equation \ref{eqn:eval_point} with the sum running from $1$ to $N_{cell}$, and multiplying the density estimate by the cell volume.\\
\indent A simple check reveals that the total mass in our SPH particle
distribution is $514.8$ M$_{\odot}$, compared to $513.3$ M$_{\odot}$
in the adaptive grid, implying that our interpolation method is
adequate. In Figure \ref{fig:grid_allgas}, we compare column densities
maps, as seen along the three principle axes, of the SPH particle
distributions (top row) and the density grids created for {\sc
  mocassin} (bottom row). We see that the {\sc mocassin} grids appear
to reproduce the morphology of the SPH particle distribution very
well. It should be noted that, in order to facilitate visualisation, the information
contained by each subgrid and SPH particle has been averaged and mapped onto a uniform pixel map, so that the resolution of the figures shown is much lower than that actually
used in our calculations. \\
\indent To compare the H~ {\sc ii} regions generated by the two codes,
we placed sources emitting $Q_{\mathrm{H}}=10^{49}$ (ionising) photons
s$^{-1}$ at position $(0,0,0)$ in both simulation volumes. The modifications described in Section 4 were disbaled, so that the SPH
code was operating in a pure Str\"omgren volume mode. We first used extremely
metal-poor abundances for the {\sc mocassin} run. The general H~{\sc
  ii} region morphologies predicted by the two codes are shown in
Figure \ref{fig:grid_hii} and are in good agreement, although a closer
look will reveal low density ionised protrusions in the SPH maps that
are absent or much less accentuated in the {\sc mocassin} maps. In
terms of total ionised mass fractions predicted, however, the
morphological discrepancies seem to have little significance, with
total ionised masses being predicted by the SPH algorithm and the {\sc
  mocassin} code of $20.9$M$_{\odot}$ and $20.5$M$_{\odot}$
respectively. The extra features appearing in the SPH maps in fact
tend to be of much lower densities. The qualitative morphological 
differences but quantitative agreement may be caused by the inherently 
poorer resolution of low--density regions by the SPH code, or it may
be an artifact of the interpolation of the Lagrangian SPH 
mass distribution onto a set of Eulerian grids. However it is more
probable instead that temperature effects may be the dominant cause,
as described in the next section. \\
\subsection{Temperature-dependance of the recombination coefficient} 
\begin{figure}
\includegraphics[width=0.45\textwidth]{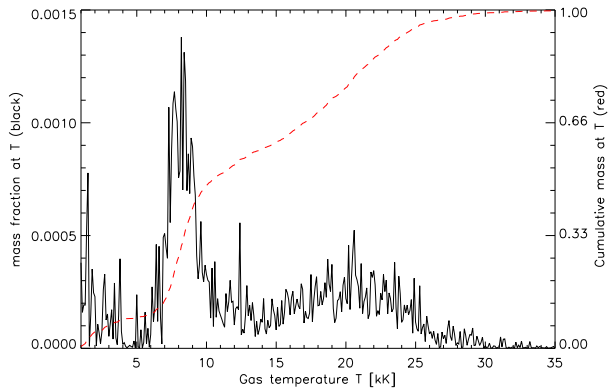}
\caption{Mass fraction (black solid line) and cumulative mass
fraction (red dashed line) of material at temperature T in the ionised
region.}
\label{fig:mocTemps}
\end{figure}

\indent The isothermal nature of the SPH
ionisation algorithm introduces a certain level of uncertainty in the
calculation of the ionised region due to the temperature dependence of
the recombination coefficient. The {\sc mocassin} model presented here, where
the gas temperatures are calculated self-consistently 
with the ionisation structure, shows that the ionised region of such a
complex density field is far from being isothermal. The
temperature structure of the ionised region is illustrated in Figure~\ref{fig:mocTemps} as the
histogram of mass fraction (black solid line) and the cumulative mass
fraction (red dashed line) of material at a given temperature.
It is clear that the success of the SPH algorithm to calculate the
correct ionised mass of the grid relies on the careful choice of the
value of the recombination coefficient, and implied gas temperature. 
For a density inhomogeneous pure hydrogen nebula of arbitrary
geometry, the total ionised volume is obtained from the solution of
\begin{equation}
\int_V{N_pN_e\alpha(T){\cdot}dV} = Q_{\rm{H}}
\end{equation}
\noindent where, for almost complete ionisation, $N_{\rm
  p}$\,=\,$N_{\rm e}$\,$\approx$\,$N_{\rm H}$, the hydrogen number
  density. The total ionised mass of a region under such assumptions,
  therefore, strongly depends on the recombination rate averaged
  temperature, defined as, 
\begin{equation}
<T[N_H^2]> = \frac{\int_V{N_H^2\,T{\cdot}dV}} {\int_V{N_H^2{\cdot}dV}}
\end{equation}
\noindent The {\sc mocassin} calculations yield $<T[N_H^2]>$~=~9120~K,
 which is very close to the value 8950~K implied by the choice of
 $\alpha_B~=~3.0{\cdot}10^{-13} cm^3 s^{-1}$ in the SPH calculation. The
 proximity of these two temperatures allowed for the excellent
 agreement in the global ionised masses calculated by the two codes.
As anticipated in the previous section, however, the detailed ionised
 volume shapes do show some differences suggesting that the
 temperature structure along a given line of sight may deviate from
that implied in the SPH.

The temperature at any point in the ionised region is
determined by the balance between the heating, dominated by the mean
energy of the photoelectrons absorbed by H and He, and cooling, which
when metals are present, is dominated by collisionally excited emission lines (particularly
infra-red fine-structure lines of $[$O~{\sc iii}$]$). In the absence
of metals, the cooling is dominated by collisionally excited
Ly$\alpha$, which is much less efficient at removing energy from the
system, therefore metal-rich regions show
lower overall temperatures and steeper temperature gradients than
their pure-H counterparts. 
Metallicity will therefore influence the temperature structure, and
consequently the ionised mass of a region. This can
be taken into account in the SPH ionisation algorithm by
choosing a different alpha value appropriate for each case.
 A good estimate of the alpha value can be obtained by running
 snapshots of the evolving density structure calculated by the SPH code at some
representative ages through a 3D photoionisation code, such as
{\sc mocassin}, to self-consistently calculate the temperature
and ionisation structure and thus obtain the recombination rate average
temperature for the grid. The appropriate alpha value for the
SPH ionisation code will therefore be that calculated at the recombination rate
averaged temperature found. \\
We should finally note that the range of gas temperature obtained
throughout the ionised region may indeed affect the hydrodynamical
evolution of the system (the
SPH ionisation algorithm assigns the same gas temperature to all
ionised particles). We defer the investigation of these potentially
important effects to future work.\\
\subsection{Effects of dust}
None of the calculations presented here include the effects of dust, which would in reality also act as a sink of ionising photons -- for example, \cite{1989ApJS...69..831W} suggest that up to $90\%$ of the stellar flux in an HII region may be absorbed by dust. It is not clear what effect dust would have on the morphology of complex HII regions such as those studied here. If the dust distribution closely follows that of the gas, the effect would essentially be to shrink the HII region but to preserve its shape. This effect could be be approximately modelled by modifiying the recombination coefficients used in the codes. However, if the dust and gas are not strongly correlated, two different radiative transfer problems must be solved simultaneously. In this the coupling of the radiative transfer solutions to the hydrodynamic evolution of the system in question is much more complicated, but we do not attempt to address this problem here.
\section{Conclusions}
We presented the results of tests of a new fast algorithm for simulating ionising radiation from point sources in the context of smoothed particle hydrodynamics (SPH) simulations. The method we use is essentially a Str\"omgren volume technique. We use a method similar to that presented in \cite{2000MNRAS.315..713K} to estimate the density profiles along lines of sight to individual SPH particles before performing a Str\"omgren integral to calculate how much flux, if any, reaches that particle. We improved on the simple Str\"omgren volume method by including in the flux calculation the flux required to ionise any neutral material lying on the path to the target particle. This modification is relevant in problems were the dynamical timescales can be as short as the timescale required to ionise the gas, which can be the case in accretion flows. We also developed a method to allow ionised particles which are somehow deprived of their photon supply to recombine and cool using an optically--thin cooling curve. This modification is also important in dynamical applications where gas can move out of an H~
{\sc ii} region, or be shadowed by dense material intervening between it and the radiation source.\\
\indent We conducted simple one--dimensional tests in which we showed that the algorithm was able to reproduce the ionisation front's R--type phase, in which it approaches the Str\"omgren radius at very high velocity, and the D--type phase in which the H~{\sc ii} region expands thermally at supersonic but ever--decreasing speed. We showed that the SPH code is able to reproduce the D--type expansion phase well even when the number of particles inside the H~{\sc ii} region is well below the canonical SPH resolution limit.\\
\indent Having demonstrated that the algorithm is able to reproduce
the analytical solutions to simple problems, we compared the results
generated in a highly inhomogeneous and anisotropic gas distribution
to the output from {\sc mocassin}, a much more sophisticated Monte
Carlo photoionisation and radiative transfer code. We found that, for
an appropriate choice of the recombination coefficient $\alpha$, the two codes agreed very well, to within $2$ percent, on the quantity of gas that should be ionised, and agreed reasonably well on the morphology of the H~{\sc ii} region. We conclude that the algorithm is adequate for modelling photoionisation in SPH simulations of star formation with very complex gas distributions.\\

{\bf Acknowledgments: } JED is supported by the Wenner--Gren Foundation, and also gratefully acknowledges many usueful discussions with Matthew Bate while in the process of devising the algorithm presented here.\\
BE was partially supported by {\it Chandra} grants GO6-7008X and
GO6-7009X. \\

\bibliography{myrefs2}

\label{lastpage}

\end{document}